\documentclass[journal]{IEEEtran}
\usepackage{CJK}
\usepackage{amssymb}
\usepackage{amsmath}
\usepackage{graphicx, subfigure}
\usepackage{url, cite}
\usepackage{verbatim}
\usepackage{booktabs}
\usepackage{multirow}
\usepackage{bigstrut}
\usepackage{booktabs}
\usepackage{amsmath,cite,graphicx,times,subfigure,url,verbatim}
\usepackage{algorithmic}
\usepackage{algorithm}
\usepackage{epstopdf}

\newtheorem{subsec:coding}{subsec:coding}

\usepackage{color}

\begin{document}

\title{Smart-Edge-CoCaCo: AI-Enabled Smart Edge with Joint Computation, Caching, and Communication in Heterogeneous IoT}

\author{
Yixue~Hao,~Yiming~Miao,Yuanwen~Tian,~Long~Hu,~M. Shamim~Hossain,~Ghulam~Muhammad,~Syed~Umar~Amin
\thanks{Y. Hao, Y. Miao Y. Tian and L. Hu are with Huazhong University of Science and Technology. (E-mail: yixuehao@hust.edu.cn, yimingmiao@hust.edu.cn, yuanwentian@hust.edu.cn, hulong@hust.edu.cn)}
\thanks{M.S. Hossain is with the department of Software Engineering, College of Computer and Information Sciences, King Saud University. (E-mail: mshossain@ksu.edu.sa)}
\thanks{G. Muhammad and S.U. Amin are with the Department of Computer Engineering, College of Computer and Information Sciences, King Saud University. (Email: ghulam@ksu.edu.sa, samin@ksu.edu.sa)}
\thanks{Long Hu is the corresponding author }
}
\maketitle

\begin{abstract}
The development of mobile communication technology, hardware, distributed computing, and artificial intelligence (AI) technology has promoted the application of edge computing in the field of heterogeneous Internet of Things (IoT). In order to overcome the defects of the traditional cloud computing model in the era of big data. In this article, we first propose a new AI-enabled smart edge with heterogeneous IoT architecture which combines edge computing, caching, and communication. Then, we propose the Smart-Edge-CoCaCo algorithm. To minimize total delay and confirm the computation offloading decision, Smart-Edge-CoCaCo uses joint optimization of the wireless communication model, the collaborative filter caching model in edge cloud, and the computation offloading model. Finally, we built an emotion interaction testbed to perform computational delay experiments in real environments. The experiment results showed that the computation delay of the Smart-Edge-CoCaCo algorithm is lower than that of traditional cloud computing model with the increase of computing task data and the number of concurrent users.
\end{abstract}


\markboth{Under review: IEEE Network, VOL. XX, NO. YY, MONTH 20XX}{}

\section{Introduction}\label{sec:introduction}

The development of Internet of Things (IoT), wireless communication technologies, and computing science has brought about tremendous changes in people-to-people communication~\cite{liang3}. From 1G and 4G to the rapidly evolving 5G mobile communication technology, IoT applications further realize the interconnections between things and things as well as people and things, based on these increasingly powerful modern communication technologies~\cite{qian1}. IoT becomes a bridge for communication between humans and physical world. Due to the extensive fields and disciplines involved, the development of IoT and mobile ubiquitous technology has transformed the shape of technological innovation in recent years. The nature of IoT applications has led to a complication of wireless communication systems and overlapping coverage of the IoT structure, which has formed a unique heterogeneous IoT for the modern physical information world.

With the current development of mobile communications technology, the 5G application scenarios defined by 3GPP provide three computing modes: Enhanced Mobile Broadband (eMBB), Massive Machine Type of Communication (mMTC), and Ultra Reliable \& Low Latency Communication (uRLLC)~\cite{CaaS}. However, the heterogeneity of IoT applications and devices lead to higher requirements for low cost, energy efficient, wide coverage, low delay, and reliable communication. To satisfy complex and diverse user computing tasks and content requests, operators adopt cloud computing technology to make up for the limitations of storage and computing in devices~\cite{Bastug2017toward}.

Nowadays, heterogeneous IoT devices have an increasing number of functions. For example, smart phones can implement many functions that were previously limited to computers and televisions. This leads to cloud computing being further approached to users~\cite{INFOCOM-2013}. However, due to the heterogeneity of the IoT, the massive data and the extremely high latency of the IoT to the cloud, the concurrent access of IoT devices has further intensified the contradiction between high bandwidth requirements and the lack of spectrum resources. The traditional single cloud access mode cannot meet the quality of experience (QoE). As a result, cloud computing has extended as edge computing of the IoT in the process of advancement~\cite{patel2014mobile}. However, current edge computing still faces challenges. The services in heterogeneous IoT are characterized by intensive and high concurrency, especially in dealing with delay-sensitive applications (such as virtual reality, emotion recognition~\cite{wearble}, etc).

The deployment of Artificial Intelligence (AI) technology in the remote cloud and edge cloud is an effective way for implementing intelligent services~\cite{liang1}. We can deploy intelligent algorithms on the cloud or edge clouds to implement new smart IoT applications~\cite{liang2}. Researches in this area include robots, language recognition, image recognition, natural language processing, and expert systems. Since the birth of AI, theory and technology have become increasingly mature, and the field of AI applications has also continued to expand. For example, AI researchers perform deep processing and analysis of unstructured data through deep learning technologies, and provide more intelligent cognitive services. As a recent research hotspot, cloud computing has begun to use AI's cognitive functions to enhance QoE. In addition, the edge cloud, combined with AI technology, is also expected to promote the development of edge computing by providing distributed AI services.

Therefore, in combination with AI, cloud computing, edge computing~\cite{shi2016edge}, and heterogeneous IoT, we want to reduce the total delay to achieve the joint optimization of computation, caching and communication of edge clouds. However, The following challenges still need consideration:

\begin{itemize}
\item {\bf Content caching strategy:}  We must consider content caching strategy from the point of improving the cache hit ratio~\cite{li2017colla}. From the caching perspective of the edge computing node, when the user requests content or computing results and the service server already stores relevant data, the content or result may be directly fed back to the user. This reduces the computation delay of the edge computing node. From the communication perspective, when the edge computing node has cached the required data, the user can directly initiate a request to the edge cloud without sending a request to the remote cloud. The data traffic of the core network will be significantly reduced, which helps to avoid network congestion~\cite{chen2017green}.
\item {\bf Computation offloading strategy:} A computation can not only be processed locally, but also on the edge cloud and remote cloud~\cite{tonghierarchical}. Therefore, it is challenging to make a computation offloading decision with different nodes (where to offload). Furthermore, even though edge computing nodes have more computing resources than local devices, computation bottlenecks can still occur when dealing with large numbers of concurrent computing tasks or very complex computing~\cite{mach2017mobile}. To avoid a sharp drop in edge computing speed, we tend to offload complex computing to remote cloud execution to optimize computation latency.
\end{itemize}

Thus, in this article, we propose a new AI-enabled smart edge with joint computation, caching, and communication in heterogeneous IoT (Smart-Edge-CoCaCo). The Smart-Edge-CoCaCo establish new modes of computation, caching and communication for these emerging technologies and applications to improve the efficiency of user-service processing, and reduce computation and communication delays. The main contributions of this article include:

\begin{itemize}
  \item We propose the AI-enabled smart edge with heterogeneous IoT architecture for the network congestion caused by the traditional cloud computing model. A new heterogeneous architecture for IoT that combines edge computing, caching, and communication is introduced to improve the hardware infrastructure in order to reduce delay and improve user QoE.
  \item We propose the Smart-Edge-CoCaCo algorithm to implement joint optimization mechanism of edge computation, caching and communication. Through the combination of wireless communication model, collaborative filtering cache model of the edge computing node, and computation offloading model, we define the minimum time delay objective to determine the computing offloaded position and the offloading delay.
  \item We conduct an experiment under the real environment on the AIWAC (Affective interaction through wearable computing and cloud technology) emotion interaction system. The experimental results showed that the average delay in the AI-enabled smart edge with heterogeneous IoT architecture with Smart-Edge-CoCaCo algorithm was shorter than traditional cloud computing.
\end{itemize}

The article is organized as follows: The AI-enabled smart edge with heterogeneous IoT architecture is presented in Section~\ref{sec:system}. The algorithm and model of Smart-Edge-CoCaco are described in Section~\ref{sec:caching}. The obtained experimental results and related discussions are given in Section~\ref{sec.performance}. Finally, Section~\ref{sec.conclusion} concludes the article.

\section{AI-Enabled Smart Edge with Heterogeneous IoT Architecture} \label{sec:system}

\begin{figure}
\centering
\includegraphics[width=3.5in]{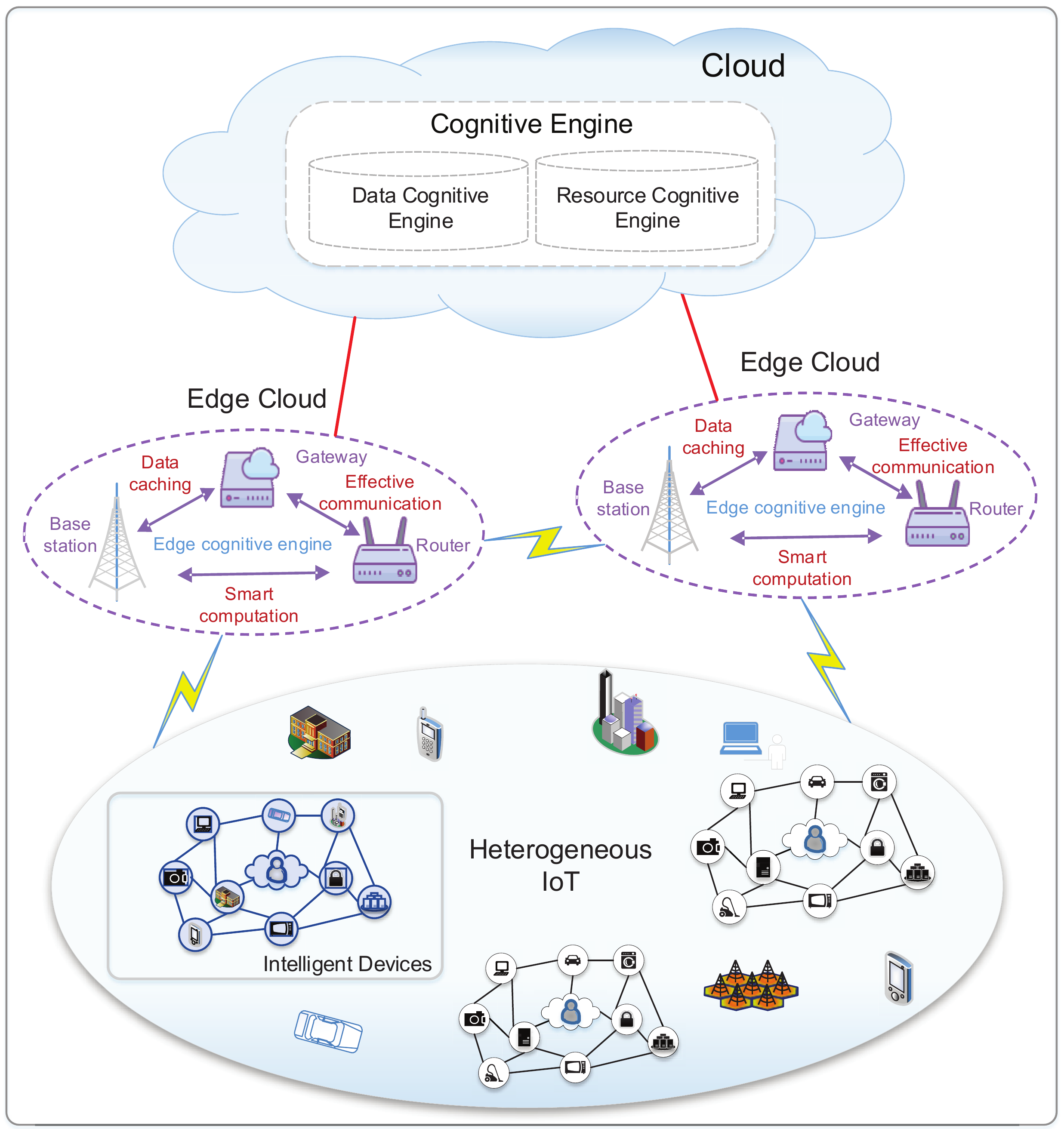}
\caption{AI-enabled smart edge with heterogeneous IoT architecture}
\label{fig1}
\end{figure}

To realize the humanization, intelligence and automation of heterogeneous IoT services, a new network structure must be designed to address the current weaknesses of the cloud-central network (traditional cloud computing mode). Thus, in this section, we propose the architecture of AI-enabled smart edge with heterogeneous IoT, as shown in Fig.~\ref{fig1}. This new architecture includes heterogeneous IoT, edge cloud and remote cloud. Through the introduction of edge computing and AI technology, we can dynamically monitor and adjust the configuration of network resources, and realize real-time data collection of IoT, efficient processing of computation, and intelligent interaction of smart applications.

\subsection{Heterogeneous IoT}

Users are the group most direct contact to the smart services provided by operators in heterogeneous IoT. They can receive feedback on the QoE based on the actual effect. The applications and services provided by these smart terminals cover almost all the environments in which people live, including smart phones, wearable or portable devices, smart homes, smart transportation, smart cities, etc. The AI-enabled smart edge with heterogeneous IoT architecture we designed mainly uses the delay-sensitive application service as the most important research subject to ensure the dynamic adjustment of network communication resources and efficient caching and computing. The most important function of the intelligent terminal is to collect user's multi-modal data. For example, a mobile phone can collect user's voice, images, action videos, etc., and request computing resources or content of interest through relevant operations to the local, edge cloud or remote cloud. The server can acquire the user's audiovisual, physiological, and emotional data resources, and perform data mining and data processing through the cognitive engine. Eventually, the computation results are obtained. Based on this, it pushes the requested content, feedback computing, and subsequent interactions to intelligent terminal.

\subsection{Intelligent Devices}

Intelligent devices are consist of a variety of multimedia sensors. Sensor is the key equipment for collecting and transmitting user data. Human sensor is responsible for collecting various human physiological information, including portable device, wearable device, etc. Dynamic sensor can sense human motion and audiovisual information, including voice sensors, cameras, gyroscopes, accelerations, GPS, etc. Edge clouds or remote clouds can infer people's intent through complex model analysis by synthesizing the perception of people's surroundings and own information.

\subsection{Edge Cloud}

The edge computing node is an important part of the edge cloud and supports the edge caching and computing functions. The edge computing nodes can be gateways, routers, switches, service servers, wireless access points, micro base stations, etc. It migrates the computing power of the remote cloud to the edge of the IoT, buffers the big data flow generated by smart devices, and reliefs network congestion during transmission. With the help of edge computing nodes, cloud computing no longer assumes all the computing tasks, but is instead more focused on investing computing resources in high-precision computing to provide optimal computing services. Edge cloud computing capacity is weaker than cloud computing capacity. However, its communication delay is much lower than the local-to-cloud. Under normal circumstances, the equipment of the IoT is connected to the edge computing nodes through wireless network, and the communication delay is greatly reduced so it can meet the demand of delay-sensitive computation. We deploy the Smart-Edge-CoCaCo algorithm on the edge cloud to achieve efficient integration and optimization of computation, caching, and communication.

\subsection{Cognitive Engine}

The cognitive engine is the brain of AI-enabled Smart Edge with Heterogeneous IoT. The remote cloud can provide high-precision computing, it deploys high-performance artificial intelligence algorithms and stores a large amount of user history analysis data and information. In addition to deploying cognitive engines in the cloud, we also deploy simple cognitive engines (intelligent algorithms for content caching and computation offloading) in the edge clouds to improve the performance of edge cloud. Cognitive engines is consist of data cognitive engine and resource cognitive engine. Data cognition engine includes AI algorithms about data mining, machine learning, deep learning, etc. Resource cognition engine includes SDN (Software defined network), NFV (Network function virtualization), SON (Self-organizing network), network slicing technologies.~\cite{Long2018} give a detailed description of these two engines.

\section{Smart-Edge-CoCaCo Algorithm}  \label{sec:caching}

In this section, we will propose a Smart-Edge-CoCaCo algorithm to improve communication efficiency and cache hit ratio, so as to achieve efficient computation offloading. On the other hand, for different computation complexity and data volume, our method can reasonably select the computation offloading location so as to balance the network load, relieve network congestion, and reduce the delay of the computation offloading.

However, there are some challenges for Smart-Edge-CoCaCo: The decision of which computation task should be processed locally and which tasks should be processed on the edge cloud or remote cloud. The edge caching and offloading aim to reduce the total delay of the computation. Therefore, we need to consider the computation, caching, and communication of the edge computing node comprehensively, optimize the computation offloading strategy, determine where to offload, and reduce the total delay of the computation.

\subsection{Wireless Communication Model}

In a cloud-based network, the user's computation task or content request offload typically through data transmission over the uplink of the wireless network. After preprocessing, a computation, or a request content, goes through cloud computing (data analysis and processing), and then the cloud returns the computation result to the user through the downlink. Here, for computation offloading delay, we do not consider downlink delay because the downlink bandwidth resources are adequate in the case of the same calculation results and the computation result packet is quite small. Here, we introduce the general wireless communication model and define the uplink data rate when offloading the computation to the edge cloud. Let $h$ and $p$ respectively represent channel power gain and transmission power of devices. Then, the offload data rate of computation $Q_{i}$ can be defined as $r=B\log_{2}\left({1+\frac{ph^{2}}{\sigma^{2}}}\right)$, where $\sigma^{2}$ represents the noise power, and $B$ represents the channel bandwidth.

\begin{figure}
\centering
\includegraphics[width=3.5in]{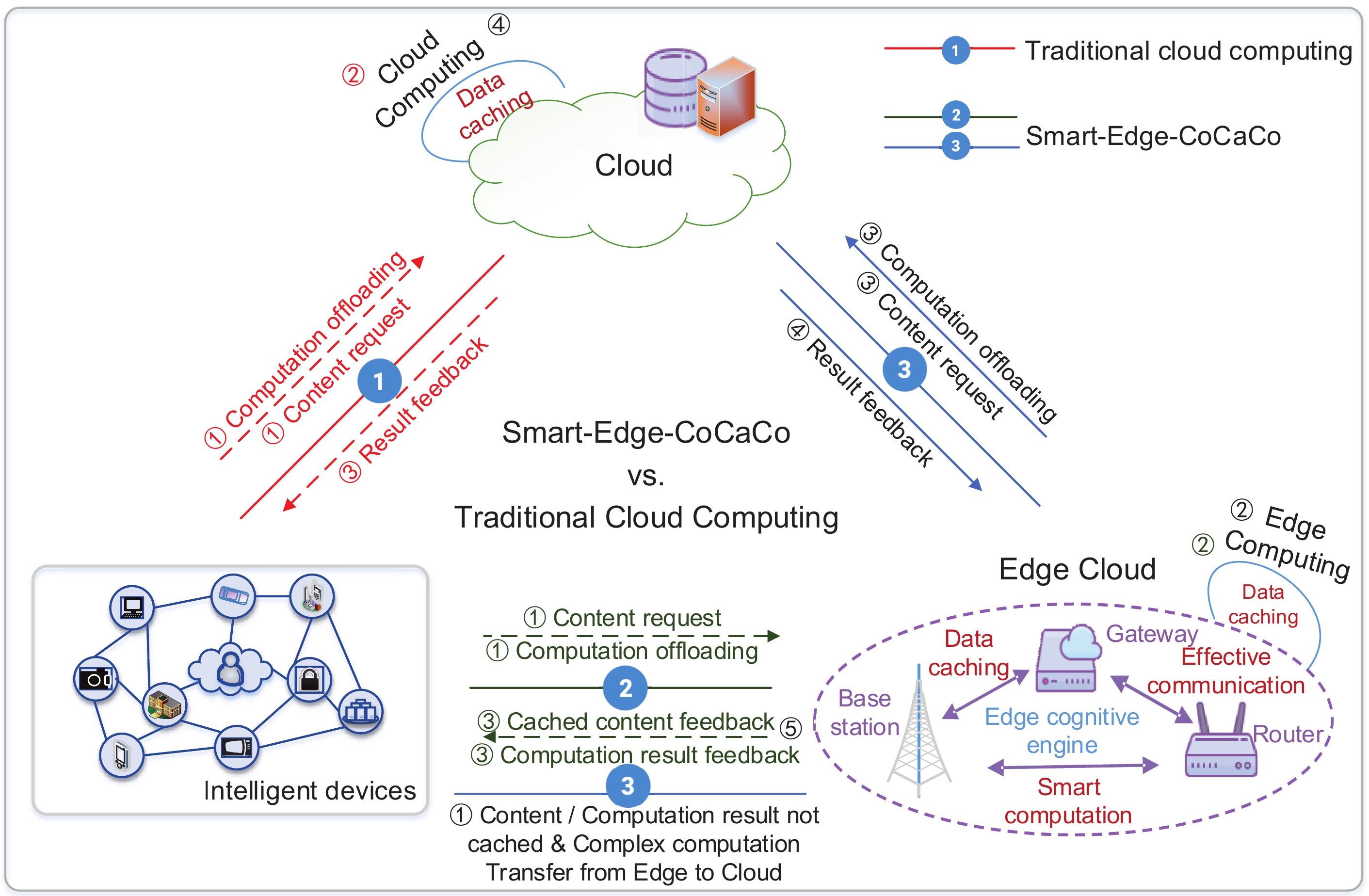}
\caption{Computation offloading strategies of Smart-Edge-Cocaco and traditional cloud computing}
\label{fig2}
\end{figure}

\subsection{Collaborative Filtering Caching Model}

As the mobile communication technology era is changing from 4G to 5G, users and devices in the IoT are usually distributed in communities. Terminal devices in the same community share one or more edge computing nodes. For an edge computing node, there are some challenges in computation and caching capabilities: the heterogeneity of IoT applications, the high concurrency of user requests, and the density of delay-sensitive intelligent services. If an edge computing node can find the data originally saved in the edge from the storage unit in time and feedback to the user when the user initiates the content request or the offload instruction, such solution can not only reduce network congestion, decrease computing and communication delay, but also bring about better user QoE.

However, what is cached is the main issue of edge computing caching strategies. First, we need to make sure what content or computation results are popular, that is, the similarity of the request content or the computation result for different users. Therefore, we use a collaborative filtering cache model to cache content or computation with similar popularity. The collaborative filtering cache model can also be used to predict what content is usually requested, contribute to dynamically adjust computation resources, and reduce computation delays. The domain-based collaborative filtering caching algorithm can be further subdivided into user-based collaborative filtering and content-based collaborative filtering.

Both user-based and content-based caching have their scope of application. The former is suitable for pushing content according to multi-user history interest and making active content pushing more personalized. The latter tends to filter the cache based on multi-user concurrency requested content or computation offloading and can retrieve and feedback the result of the computing in a timely manner, which is more social. In general, we tend to use user-based collaborative filtering caching strategies in remote cloud. The reason is that the permanent storage of user historical data, and powerful data analysis and computation capabilities. However, this is not suitable for edge computing nodes. On the contrary, we deploy a content-based collaborative filtering caching algorithm on the edge cloud.

The user's requested content or offloaded computation is represented as a vector, and the cosine similarity represents the angle between the two vectors. The smaller the angle is, the more similar the user needs. Here, we use the Euclidean Distance to measure the similarity between the requested content / offloaded computation and cached content / result. A similarity computing function for offloaded data can be defined as $sim(u,v_{j})=(\sum_{i=1}^{I}\vert{u_{i}-v_{ji}}\vert^{2})^\frac{1}{2}$, 
where $u$ represents a user-initiated request or uninstalled computation. $I$ represents the total number of elements in the vector. $v_{j}$ represents the current comparison of the edge cloud cache, where $j\in{N}$, $N$ represents the total amount of cached content. Therefore, we compare the user-initiated request or offloaded computation with all cached content.

After obtaining the similarity result of the requested content or computation, the highest value of the similarity can be defined as  Eq.(\ref{eq:3}),

\begin{equation}\label{eq:3}
\overline{r_{u}}=\lfloor\max({sim(u,v_{j})})+\frac{1}{2}\rfloor
\end{equation}
where $r_{u}$ represents the cache hit ratio for the user's request. In order to provide a better service experience and achieve computation efficiency of edge computation offloading, we round off the cache hit ratio, i.e., the final cache hit ratio $\overline{r_{u}}\in[0, 1]$. Therefore, if the edge cloud does not cache the related content or the computing result, the edge computing node needs to immediately perform the computation, and the total delay will include the computing delay and communication delay. Otherwise, the computing delay can be saved.

\subsection{Smart-Edge-CoCaCo Computation Offloading Model}

In the Smart-Edge-CoCaCo algorithm, we introduce the edge computing and caching technology to perform dense application computing on the edge of the IoT while guarantee the low delay. As shown in Fig.~\ref{fig2}, there are 3 modes for computation offloading. Our Smart-Edge-CoCaCo algorithm includes two modes about Edge computing (procedure 2) and Edge-Cloud collaborative computing (procedure 3), while traditional cloud computing only follow up procedure 1. This is why we have introduced edge computing. We should determine where to offload based on the amount of data generated by the user or the complexity of the computation, that is, whether or not to directly offload computation to the cloud.

\begin{itemize}
\item Traditional cloud computing: As shown in procedure 1 in Fig.~\ref{fig2}, a large number of users and their smart devices will generate massive amounts of multi-modal data every day. The concurrency and heterogeneity content requests received by the cloud have brought tremendous pressure on network spectrum resources and bandwidth utilization. The network congestion and communication and computing delay increases dramatically if there is no intelligent algorithm to implement regulation.
\item Edge computing: As shown in procedure 2 in Fig.~\ref{fig2}, if the content or computation result requested by the user is stored in the edge cloud, the service server at the edge of the IoT can directly feedback the content or result to users, which can effectively reduce the computation delay. If the edge cloud does not cache relevant content, but the computation complexity of the user's offloading is within the computation capacity of the edge computing node, it can also perform the computation and feedback the results to the user.
\item Edge-cloud collaborative computing: As shown in procedure 3 in Fig.~\ref{fig2}, if the edge cloud does not cache related content and the computation complexity exceeds the capacity of the edge computing node, the edge cloud will offload computation to the cloud. The cloud will feed back the cached content or computation results to the user via the edge cloud.
\end{itemize}

The Smart-Edge-CoCaCo communication, caching, and computation joint model, that combines the above discussion of edge computing offloading with the collaborative filter caching strategy, can be defined as Eq.(\ref{eq:4}),

\begin{equation}\label{eq:4}
\begin{cases}
T_{LtoC}=\frac{R}{r_{E-up}}+\frac{D}{P_{Cloud}}+\frac{F}{r_{E-down}}+\alpha_{EtoC}\\ T_{LtoE}=\frac{R}{r_{E-up}}+\frac{D}{P_{Edge}}\ast{(1-\overline{r_{u}})}+\frac{F}{r_{E-down}}\\
\end{cases}
\end{equation}
where, $T_{LtoC}$ represents the total delay of the computation offloaded from local to cloud through edge and receiving feedback of the results (edge-cloud collaborative computing in procedure 3). $T_{LtoE}$ represents the total delay for the computation offloaded to the edge cloud and receive feedback of the results (edge computing in procedure 2). $R$ represents the amount of data sent by the user, that is, the amount of offloaded data. $D$ represents the amount of data or complexity of the computation. $F$ represents the amount of feedback computation results or content data, which is usually every small. $P_{Cloud}$ represents the processing power of the cloud. $P_{Edge}$ represents the processing power of the edge computing node. $r_{E-up}$ represents the offload data rate from local to the edge cloud, which is computed by the wireless communication model. $r_{E-down}$ represents the download data rate from the edge cloud to the local computer which is very large. The download delay can be ignored due to the small result data and large downlink data rate. $\alpha_{EtoC}$ represents the optical fiber transmission delay from edge to cloud which is very small. $\overline{r_{u}}$ represents the cache probability of computation/requests in the edge cloud, computed by Eq.(\ref{eq:3}). $\overline{r_{u}}=1$ means the edge cloud cached related content and the computation delay in edge is saved. If $\overline{r_{u}}=0$, we should compare the the total delay of edge-cloud collaborative computing and edge computing, so as to determine where to offload. After getting the total delay of above two modes, we will choose the offload location based on the smaller time delay between the two, that is, the objective function of this model is $min(T_{LtoC}, T_{LtoE})$. When $T_{LtoE}$ is smaller, the users will offload their computation to edge cloud. If $T_{LtoC}$ is smaller, the edge cloud will transmit the complex computation to the remote cloud to get better performance.

\section{Performance Evaluation} \label{sec.performance}

In this section, we discusses the performance experiments on the Smart-Edge-CoCaCo algorithm and the traditional cloud computing mode under real-world conditions. We use the AIWAC emotion recognition system~\cite{Long2018} to perform computation offloading experiments for facial expression recognition applications. Specifically, the proposed architecture uses smart phones as intelligent devices in real environments, the service server as the edge computing node, and the data center as the remote cloud. At the same time, deep learning algorithms are deployed at each computing node to provide facial expression recognition services. The purpose is to test the impact of the computation load and the number of concurrent users on the total delay, and compare the performance of the Smart-Edge-CoCaCo algorithm with the traditional cloud computing mode.

\begin{figure}
\centering
\includegraphics[width=3.3in]{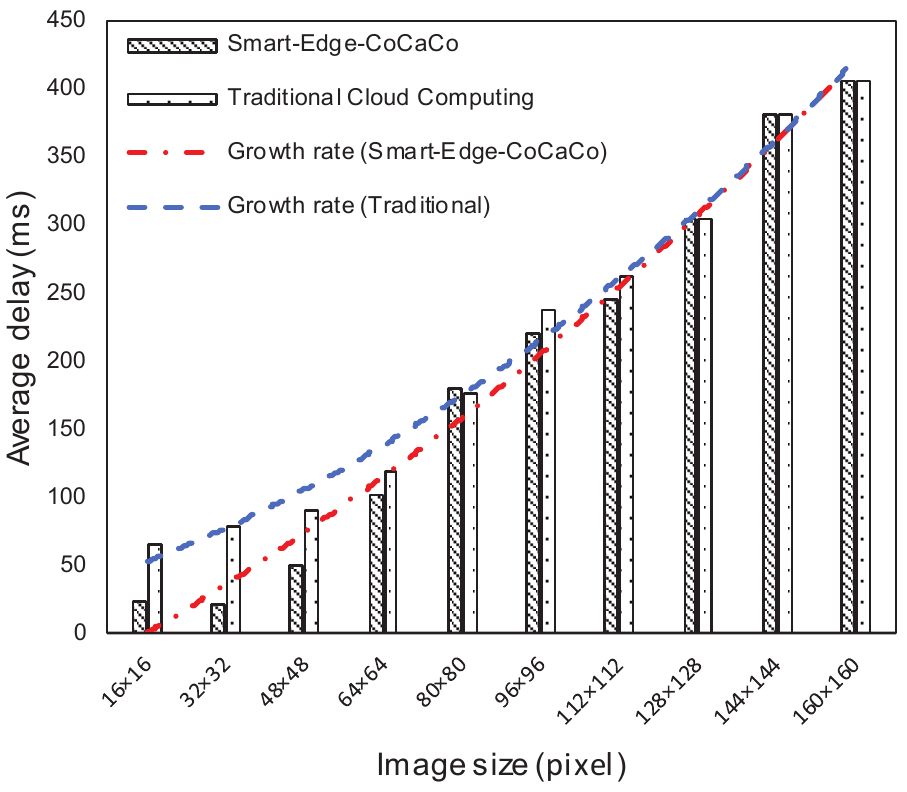}
\caption{Task computing comparison between Smart-Edge-CoCaCo and traditional cloud computing}
\label{fig3}
\end{figure}

More concretely, our experiment is testing the performance of the Smart-Edge-CoCaCo algorithm through the communication and interaction between mobile devices and edge computing nodes or cloud data centers. To test the computation bottlenecks in the edge computing nodes and cloud data centers in the real environment, we performed emotion recognition with different image sizes in two computation offloading modes (Smart-Edge-CoCaCo algorithm and traditional cloud computing). The total delay of the computation is shown in Fig.~\ref{fig3}.

In the first experiment, we offload different computations to test the average delay of two modes. The sizes of facial expression image we used are $Image_{width}\in[16,32,48,64,80,96,112,128,144,160]$, which also shows that the computation complexity or amount of data in emotion recognition applications increases gradually. As shown in Fig.~\ref{fig3}, the overall delay of the Smart-Edge-CoCaCo algorithm is lower than that of the traditional cloud computing. This is because the direct communication between the local device and the cloud data center causes the core network load to be too large. Although cloud data center has stronger processing power as well as stronger computation and caching capabilities, its total delay was still higher than that of our proposed algorithm due to the long communication delays on the computation offloading. The Smart-Edge CoCaCo algorithm considers the computation delay and communication delay, and the computation offloading mode is more flexible and can effectively reduce the delay. Besides, when the computation complexity was low, the system tended to offload computation to the edge cloud because the processing power of the edge computing nodes met the requirements and the communication delay was low. When the computation complexity was high, the system tended to offload computing to the cloud, that is, the total delay was the same as traditional cloud computing. Therefore, the growth rate of the two tends to be consistent in the end.

\begin{figure}
\centering
\includegraphics[width=3.5in]{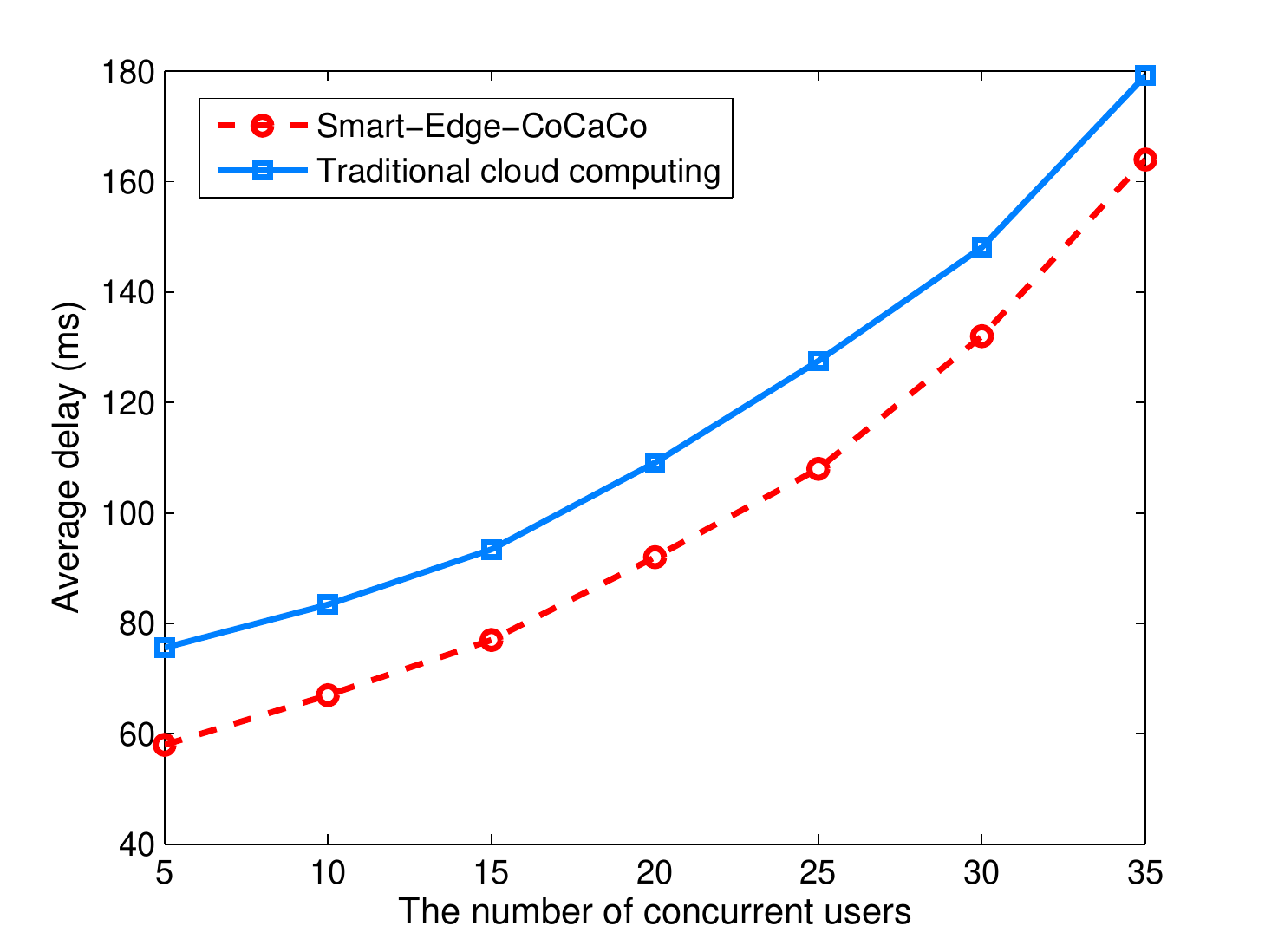}
\caption{Concurrent computing comparison between Smart-Edge-CoCaCo and traditional cloud computing}
\label{fig4}
\end{figure}

Then, we tested the average delay of two modes with increasing concurrent number of users (each user conduct offloading with same computation complexity), as shown in Fig.~\ref{fig4}. We assume that all user-initiated sentiment analysis requests are consistent and the number of users is gradually increasing to visualize the difference. Specifically, the image size $Image_{size}=16\ast16$, the number of users $Num_{user}\in[5, 10,15,20,25,30,35]$. We found that, in this kind of facial expression recognition application, the average delay of traditional cloud computing mode increased exponentially with the number of concurrent users as the user number has a greater impact on the occupancy rate of bandwidth resource. However, our Smart-Edge-CoCaCo algorithm performs better. The average delay of Smart-Edge-CoCaCo is always lower than traditional mode due to the collaborative filtering and computation offloading mechanism. Some computation will be offloaded to the near edge cloud depending on the network conditions and edge caching. Moreover, if some user request same/similar content which was already cached in the edge cloud, the delay will be greatly reduced. From the trend of delay curves, as the number of concurrent increases, the gap between these two modes will be narrowed due to the network congestion.

\section{Conclusion}\label{sec.conclusion}

In this article, we proposed a new AI-enabled smart edge with heterogeneous IoT architecture to address the disadvantages of traditional cloud computing mode in terms of communication delay and network load. Then, we proposed the Smart-Edge-CoCaCo algorithm for the joint optimization mechanism of computation, caching, and communication. From the perspectives of joint optimization of the wireless communication model, the collaborative filter caching model of the edge cloud, and the computation offloading model, the principle of minimization of delay is specified, and the computation offloading strategy is implemented and regulated. Finally, in a real experiment of the AIWAC affective interaction system, the average delay of the Smart Edge-CoCaCo algorithm was lower than the traditional cloud computing mode with the increasing computation and concurrent users.

In the future, we will conduct more research on the caching of the edge computing nodes and offloading model. Moreover, we will do additional comparative verification experiments on cache hit ratio, coverage, and offloading delay to optimize network architecture and algorithm performance, and provide theoretical and experimental references for the joint optimization of computing, caching, and offloading of edge clouds in heterogeneous IoT.

\section*{Acknowledgement}

The authors extend their appreciation to the Deanship of Scientific Research at King Saud University, Riyadh, Saudi Arabia for funding this work through the research group project no. RG-1436-023.

\bibliographystyle{IEEEtran}

\begin{IEEEbiographynophoto} {Yixue Hao} (yixuehao@hust.edu.cn) received the B.E. degree in Henan University, China, and his Ph.D degree in computer science from Huazhong University of Science and Technology (HUST), China, 2017. He is currently working as a postdoctoral scholar in School of Computer Science and Technology at Huazhong University of Science and Technology. His research includes 5G network, internet of things, mobile cloud computing.
\end{IEEEbiographynophoto}

\begin{IEEEbiographynophoto} {Yiming Miao} (yimingmiao@hust.edu.cn) received the B.Sc. degree in College of Computer Science and Technology from QingHai Univerisity, Xining, China in 2016. Currently, she is a Ph.D candidate in School of Computer Science and Technology at Huazhong University of Science and Technology (HUST). Her research interests include IoT sensing, healthcare big data and emotion-aware computing, etc.
\end{IEEEbiographynophoto}

\begin{IEEEbiographynophoto} {Yuanwen Tian} is currently pursuing the bachelor's degree with the School of Electrical and Electronic Engineering, Huazhong University of Science and Technology, China. He joined the Embedded and Pervasive Computing Laboratory as an outstanding undergraduate in terms of academic performance in 2016, supervised by Professor Min Chen. His current research interests include cognitive computing, distributed and cloud computing, Internet of Things, and big data analytics.
\end{IEEEbiographynophoto}

\begin{IEEEbiographynophoto} {Long Hu} (hulong@hust.edu.cn) is a lecture in School of Computer Science and Technology, Huazhong University of Science and Technology (HUST), China, Since 2017. He was a visiting student at the Department of Electrical and Computer Engineering, University of British Columbia from Aug. 2015 to Apr. 2017. His research includes the Internet of Things, Software Defined Networking, Caching, 5G, body area networks, body sensor networks and mobile cloud computing.
\end{IEEEbiographynophoto}

\begin{IEEEbiographynophoto}{M. Shamim Hossain} (mshossain@ksu.edu.sa)  is a Professor with the Department of Software Engineering, College of Computer and Information Sciences, King Saud University, Riyadh, Saudi Arabia. He has authored or co-authored more than 165 publications. He is the recipient ACM TOMM Nicolas D. Georganas Best Paper Award.  He currently serves on the Editorial Board of IEEE Multimedia. His research focuses on social media, Internet of Things (IoT), cloud and multimedia for healthcare, and smart health.
\end{IEEEbiographynophoto}
\begin{IEEEbiographynophoto} {Ghulam Muhammad} is a professor in the Computer Engineering Department, College of Computer and Information
Sciences, King Saud University. He received his Ph.D. degree
in 2006 from the Department of Electronic and Information
Engineering at Toyohashi University of Technology, Japan. His
research interests include speech and image signal processing,
multimedia forensics, cloud computing, and healthcare. He has
authored and co-authored many journal and conference publications.
He owns a U.S. patent.
\end{IEEEbiographynophoto}

\begin{IEEEbiographynophoto} {Syed Umar Amin} is a Ph.D. researcher in the Department of Computer Engineering, College of Computer and Information Sciences at King Saud University. He received his Master¡¯s
degree in computer engineering from Integral University, India,
in 2013. His research interests include deep learning, EEG analysis
and classification, biologically inspired artificial intelligence,
and data mining in healthcare.
\end{IEEEbiographynophoto}

\end{document}